\documentclass[%
reprint,
superscriptaddress,
preprintnumbers,
amsmath,amssymb,
aps,
prc,
]{revtex4-2}
\usepackage[dvipdfmx]{graphicx}
\usepackage{dcolumn}
\usepackage{bm}
\usepackage[T1]{fontenc}
\usepackage{braket}
\usepackage{hyperref}
\hypersetup{pdfpagemode={UseOutlines},
bookmarksopen=true,
bookmarksopenlevel=0,
hypertexnames=false,
colorlinks=true,
citecolor=blue,
linkcolor=blue,
urlcolor=blue,
allcolors=blue,
pdfstartview={FitV},
unicode,
breaklinks=true,
}

\usepackage{etoolbox} 

\begin{document}
\preprint{NITEP 119}

\title{First application of the dispersive optical model to ($p$,$2p$) reaction analysis within the distorted-wave impulse approximation framework}

\author{K. Yoshida}%
\email{yoshida.kazuki@jaea.go.jp}
\affiliation{%
Advanced Science Research Center, Japan Atomic Energy Agency, Tokai, Ibaraki 319-1195, Japan
}%
\author{M. C. Atkinson}%
\affiliation{%
Theory Group, TRIUMF, vancouver, british columbia V6T 2A3, Canada
}
\author{K. Ogata}%
\affiliation{%
Research Center for Nuclear Physics (RCNP), Osaka University, Ibaraki 567-0047, Japan
}
\affiliation{%
Department of Physics, Osaka City University, Osaka 558-8585, Japan
}
\affiliation{%
Nambu Yoichiro Institute of Theoretical and Experimental Physics (NITEP), Osaka City University, Osaka 558-8585, Japan
}
\author{W. H. Dickhoff}%
\affiliation{%
Department of Physics, Washington University, St. Louis, Missouri 63130, USA
}

\date{\today}

\begin{abstract}
\begin{description}
\item[Background]
Both ($e$,$e'p$) and ($p$,$2p$) reactions have been performed to study the 
proton single-particle character of nuclear states with its related spectroscopic factor.
Recently, the dispersive optical model (DOM) was applied to the ($e$,$e'p$) analysis
revealing that the traditional treatment of the single-particle overlap function, distorted waves, and nonlocality must be further improved to achieve quantitative nuclear spectroscopy.
\item[Purpose]
We apply the DOM wave functions to the traditional ($p$,$2p$) analysis and investigate 
the consistency of the DOM spectroscopic 
factor that describes the ($e$,$e'p$) cross section with the result of the ($p$,$2p$) analysis.
Additionally, we make a comparison with a phenomenological single-particle wave function
and optical potential.
Uncertainty arising from a choice of $p$-$p$ interaction is also investigated.
\item[Method]
We implement the DOM wave functions to the nonrelativistic distorted wave impulse approximation (DWIA)
framework for ($p$,$2p$) reactions.
\item[Results]
DOM + DWIA analysis on $^{40}$Ca($p$,$2p$)$^{39}$K data generates a proton $0d_{3/2}$ 
spectroscopic factor of 0.560, which is meaningfully smaller than the DOM value of 0.71 shown to be consistent with the ($e$,$e'p$) analysis. 
Uncertainties arising from choices of single-particle wave function, optical potential, 
and $p$-$p$ interaction do not explain this inconsistency.
\item[Conclusions]
The inconsistency in the spectroscopic factor suggests there is urgent need for
improving the description of $p$-$p$ scattering in a nucleus 
and the resulting in-medium interaction with corresponding implications for the analysis of this reaction in inverse kinematics.
\end{description}
\end{abstract}

\maketitle


\section{Introduction}
The independent particle picture provides an excellent first characterization of the structure of a nucleus.
An important indicator of this picture is the spectroscopic factor for valence orbitals, which represents the removal probability for each nucleon orbital to a low-lying state of the system with one proton less.
The nucleon knockout reaction has been one of the best tools 
to study this aspect of nuclei.
The electron-induced proton knockout reaction, ($e$,$e'p$)~\cite{Steenhoven88,Herder88,Kramer89,Huberts90,Lapikas93,Vijay97,Kramer01,Mack2018,Mack2019}, has been considered the cleanest spectroscopic method for decades. Despite some concerns about the uncertainties associated with proton-induced proton knockout reactions, ($p$,$2p$)~\cite{Jacob66,Jacob73,Chant77,Chant83,Samanta86,Cowley91,Wakasa17,Aumann21},  a recent review~\cite{Wakasa17} established $(p,2p)$ as an complementary spectroscopic tool to ($e$,$e'p$) with about 15\% uncertainty for incident energy above 200~MeV.

As discussed in Ref.~\cite{Wakasa17}, the effect of nonlocality on the distorted waves and the bound-state wave function is considered to be a major source of the theoretical uncertainties in the description of the ($p$,$2p$) reactions. 
Usually, the effect is phenomenologically taken into account by including the Perey factor~\cite{Perey62}; the Darwin factor is used when an optical potential based on the Dirac phenomenology~\cite{Hama90,Cooper93,Cooper09} is adopted.
However, the validity of this phenomenological treatment of nonlocality has not been estimated quantitatively.
Recently, a fully nonlocal dispersive optical model (DOM) has been developed~\cite{Mahzoon:2014,Dickhoff:2017}
extending the original work by Mahaux and Sartor~\cite{Mahaux91}.
The DOM describes the nucleon scattering potential and
the binding potential that gives single-particle levels on the same footing, 
making use of a subtracted dispersion relation.
The single-particle wave function (SPWF) and its spectroscopic factor as well as the distorted waves obtained by the present DOM framework
were applied to the nonrelativistic distorted wave impulse approximation (DWIA) analysis 
of $^{40,48}$Ca($e$,$e'p$)$^{39,47}$K reactions~\cite{Mack2018,Mack2019} without any further adjustment.
It was concluded that an accurate treatment of the nonlocality as practised in the DOM is necessary to generate spectroscopic 
factors that automatically describe the ($e$,$e'p$) knockout cross sections after the DOM potential has been constrained by all available elastic scattering data (up to 200 MeV) and relevant ground-state information.

The finding in Refs.~\cite{Mack2018,Mack2019} demonstrates the importance of the implementation of the DOM wave functions to the DWIA analysis of ($p$,$2p$) processes in which three proton distorted waves are present. Therefore, the use of the DOM wave functions is expected to clarify the role of the effective interaction that is involved in the extraction of spectroscopic factors from ($p$,$2p$) data.
In this study, we report the first application of such a DWIA analysis with the DOM wave functions to $^{40}$Ca($p$,$2p$)$^{39}$K data at 200~MeV~\cite{Noro_40Ca_p2p_private}.

The organization of this paper is as follows.
In Sec.~\ref{sec_theory}, we briefly introduce the present DOM + DWIA framework.
In Sec.~\ref{sec_result},
a theoretical analysis of $^{40}$Ca($p$,$2p$)$^{39}$K data and the consistency
between DOM proton spectroscopic factors that automatically describe $(e,e'p)$ cross sections and those obtained from ($p$,$2p$) analyses are discussed.
We also compare the present framework with a conventional DWIA analysis employing 
a phenomenological SPWF and optical potentials.
Finally, a summary is given in Sec.~\ref{sec_summary}.

\section{Theoretical framework}
\label{sec_theory}
\subsection{Distorted wave impulse approximation}
In the present study, the factorized form of the nonrelativistic DWIA
with the spin degrees of freedom is employed.
The transition matrix $T$ within the distorted wave impulse approximation framework
is given by
\begin{widetext}
\begin{align}
T_{\mu_1 \mu_2 \mu_0 \mu_j}
&=
\sum_{\mu'_1 \mu'_2 \mu'_0 \mu_p}
\tilde{t}_{\mu'_1 \mu'_2 \mu'_0 \mu_p} 
\int d\bm{R}\,
\chi_{1, \mu'_1 \mu_1}^{(-)*}(\bm{R})
\chi_{2, \mu'_2 \mu_2}^{(-)*}(\bm{R})
\chi_{0, \mu'_0 \mu_0}^{(+)}(\bm{R})
e^{-i\alpha_{R}\bm{K}_0 \cdot\bm{R}}
\sum_{m}
\left(lms_p\mu_p | j\mu_j\right)
\psi^n_{ljm}(\bm{R}).
\label{eq:tmatrix}
\end{align}
\end{widetext}
The incident and two emitted protons are labeled as particle $0$--$2$,
while the bound proton in the initial state is labeled as $p$.
$\chi_{i,\mu'_i \mu_i}$ is a distorted wave of particle $i=0,1,2$
having the asymptotic (local) third component $\mu_i$ ($\mu'_i$) of its spin $s_i = 1/2$.
The outgoing and incoming boundary conditions of the distorted waves 
are denoted by superscripts $(+)$ and $(-)$, respectively.
$\bm{K}_0$ is the momentum (wavenumber) of the incident proton and 
$\alpha_{R}$ is the mass ratio of the struck particle and the target, $1/40$ in this study.
$n$ is the radial quantum number, and $l, j, m$ are the single-particle
orbital angular momentum, total angular momentum, and third component of $l$,
respectively.
$\psi^n_{ljm}$ is the SPWF normalized to unity.
$\tilde{t}_{\mu'_1 \mu'_2 \mu'_0 \mu_p}$ is the 
matrix element of the $p$-$p$ effective interaction $t_{pp}$,
\begin{align}
\tilde{t}_{\mu'_1 \mu'_2 \mu'_0 \mu_p} 
&=
\Braket{
\bm{\kappa}', \mu'_1 \mu'_2 
| t_{pp} |
\bm{\kappa}, \mu'_0 \mu_p
},
\label{eq:tpp}
\end{align}
where $\bm{\kappa}$ and $\bm{\kappa}'$ are relative momenta of two protons
in the initial and the final states, respectively.
The factorization procedure of $t_{pp}$ is explained using the local semiclassical approximation
(LSCA) and the asymptotic momentum approximation (AMA) in 
the appendix.
It should be noted that the factorized DWIA is often regarded as a result of the zero-range approximation
but $t_{pp}$ is finite-range interaction in the present study.

The triple differential cross section (TDX) with respect to the 
emitted proton energy $T_1^\mathrm{L}$ and emission angles 
$\Omega_1^\mathrm{L}$ and $\Omega_2^\mathrm{L}$
is given as
\begin{align}
\frac{d^3\sigma^{\mathrm{L}}}{dT_1^\mathrm{L} d\Omega_1^\mathrm{L} d\Omega_2^\mathrm{L}}
&=
\mathcal{Z}^n_{lj}
\mathcal{J}_\mathrm{LG} F_\mathrm{kin}
\frac{(2\pi)^4}{\hbar v_\alpha}
\frac{1}{(2s_0+1)(2j+1)}
\nonumber 
\\
&\times
\sum_{\mu_1 \mu_2 \mu_0 \mu_j} 
\left| T_{\mu_1 \mu_2 \mu_0 \mu_j} \right|^2,
\label{eq:tdx}
\end{align}
with $ \mathcal{Z}^n_{lj}$, $\mathcal{J}_\mathrm{LG}$, $F_\mathrm{kin}$, $v_\alpha$ being 
the spectroscopic factor,
the Jacobian 
from the center-of-mass frame to the Laboratory frame, kinetic factor, and
the relative velocity of the incident proton and the target, respectively.
Quantities with superscript $\mathrm{L}$ are evaluated in the laboratory frame 
while the others are in the center-of-mass frame.
See Sec. 3.1 of Ref.~\cite{Wakasa17} for details.

\subsection{Dispersive optical model}

The distorted waves and SPWF in Eq.~\eqref{eq:tmatrix} as well as the spectroscopic factor in Eq.~\eqref{eq:tdx} are calculated using the DOM.
The nonlocal DOM uses both bound and scattering data to constrain the nucleon self-energy  $\Sigma^*$ for a given nucleus.
This self-energy is a complex and nonlocal potential that unites the nuclear structure and reaction domains~\cite{Mahaux91,Mahzoon:2014}.
 The DOM was originally developed by Mahaux and Sartor~\cite{Mahaux91}, employing local real and imaginary potentials connected through dispersion relations. However, only
with the introduction of nonlocality can realistic self-energies be obtained~\cite{Mahzoon:2014,Dickhoff:2017,Atkinson:2020,Pruitt:2020,Pruitt:2020C}. 
The Dyson equation then determines the single-particle propagator, or Green's function, from which bound-state and scattering observables can be deduced.

Using the DOM self-energy, a Schr\"odinger-like equation can be generated to calculate the SPWF in Eq.~\eqref{eq:tmatrix},
\begin{align}
      \sum_\gamma\bra{\alpha}T_{\ell} + \Sigma^*_{\ell j}(E)\ket{\gamma}\psi_{\ell j}^n(\gamma) = \varepsilon_n^-\psi_{\ell j}^n(\alpha),
      \label{eq:schrodinger}
\end{align} 
where $\alpha$ and $\gamma$ are arbitrary basis variables, $\ell j$ correspond to the orbital and total angular momentum of the orbital, $n$ is the principle quantum number of the orbital, and $\bra{\alpha}T_\ell\ket{\gamma}$ is the kinetic-energy matrix element, including the centrifugal term~\cite{Mack2018}.
These wave functions correspond to overlap functions
   \begin{equation}
      \psi^n_{\ell j}(\alpha) = \bra{\Psi_n^{A-1}}a_{\alpha \ell j}\ket{\Psi_0^A}, \qquad \varepsilon_n^- = E_0^A - E_n^{A-1}.
      \label{eq:wavefunction}
   \end{equation}
Such discrete solutions to Eq.~(\ref{eq:schrodinger}) exist where there is no imaginary part of the self-energy, so near the Fermi energy. 
The normalization for these wave functions corresponds to the same spectroscopic factor of Eq.~\eqref{eq:tdx}, which is given by~\cite{Exposed!}
   \begin{equation}
      \mathcal{Z}^n_{\ell j} = \bigg(1 - \frac{\partial\Sigma_{\ell j}^*(\alpha_{qh},\alpha_{qh};E)}{\partial E}\bigg|_{\varepsilon_n^-}\bigg)^{-1},
      \label{eq:sf}
   \end{equation}
   where $\alpha_{qh}$ represents the quasihole state that solves Eq.~(\ref{eq:schrodinger}). Note that because of the presence of imaginary parts of the self-energy at other energies, there is also strength located there, and thus the spectroscopic factor will be less than 1 and also less than the occupation probability. The DOM self-energies of $^{40}$Ca and $^{48}$Ca were used in Refs.~\cite{Mack2018,Atkinson20,Mack2019} to reproduce $^{40}$Ca$(e,e'p)^{39}$K and $^{48}$Ca$(e,e'p)^{47}$K momentum distributions, respectively. The corresponding $^{40}$Ca spectroscopic factor of $\mathcal{Z}_{0d_{3/2}}^{\mathrm{DOM}}=0.71\pm 0.04$, which is consistent with $^{40}$Ca$(e,e'p)^{39}$K data, will now be used alongside the DOM SPWF and distorted waves to analyze the $^{40}$Ca$(p,2p)^{39}$K knockout reaction.

\section{Results and discussion}
\label{sec_result}
In this section, we discuss extracted spectroscopic factors from the $^{40}$Ca$(p,2p)^{39}$K reaction in comparison with the DOM result that is consistent with the $^{40}$Ca$(e,e'p)^{39}$K data.
We also address  uncertainties arising from the choice of the optical potential and the effective $p$-$p$ interaction.
The theoretical knockout cross section is calculated using the 
DWIA framework with the DOM SPWF and distorted waves.
Results using phenomenological inputs are also discussed for comparison.

The spectroscopic factor of $0d_{3/2}$ is  
extracted from the ratio of the theoretical cross section 
and the experimental data  
of the $^{40}$Ca($p$,$2p$)$^{39}$K reaction at 197~MeV.
The reaction kinematics is in a coplanar kinematics and 
the opening angles of the emitted protons are
fixed at the same angle: 
$\phi_1^\mathrm{L} = 0^\circ$, 
$\phi_2^\mathrm{L} = 180^\circ$,
and
$\theta_1^\mathrm{L} = \theta_2^\mathrm{L} = 42.0^\circ$
in the Madison convention~\cite{madison_convention}.
The kinematics of the three particles is then uniquely determined once
$T_1^\mathrm{L}$ is given.

\begin{figure}[htbp]
\centering
\includegraphics[width=0.37\textwidth]{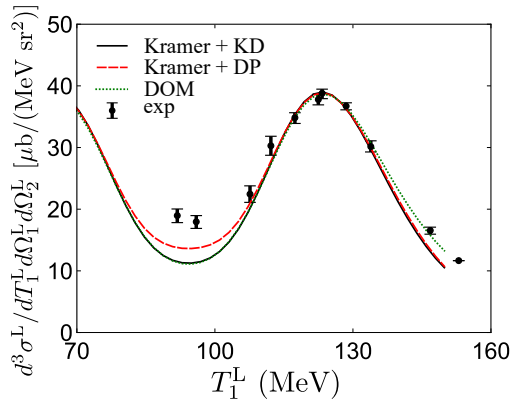}
\caption{
TDX with different optical potentials.
The solid, dashed, and dotted lines are TDXs with the Koning-Delaroche
optical potential (KD) and Dirac phenomenology (DP),
respectively.
The result with DOM ingredients is also shown as the dot-dashed line. All results reflect cross sections that are normalized with the spectroscopic factors shown in Table~\ref{tab:s-factor}.
The experimental data taken by the E258 experiment at RCNP~\cite{Noro_40Ca_p2p_private} 
are also shown.
}
\label{fig_tdx_dw}
\end{figure}

The DOM-DWIA result is compared with those of the phenomenological 
SPWF and the optical potential in Fig~\ref{fig_tdx_dw}.
For this comparison, the DOM-DWIA cross section is adjusted to the data and the DOM spectroscopic factor was not utilized.
The phenomenological SPWF 
suggested by Kramer \textit{et al.}~\cite{Kramer01}, the
Koning-Delaroche optical potential parameter set (KD)~\cite{Koning03}, 
and the
Dirac phenomenology (DP)~\cite{Hama90,Cooper93,Cooper09} are also considered.
Calculated TDXs and the experimental data are shown in Fig.~\ref{fig_tdx_dw}.
Spectroscopic factors are therefore extracted from the ratio of the present calculations and the 
experimental data taken by the E258 experiment at RCNP~\cite{Noro_40Ca_p2p_private}
by minimizing
\begin{align}
  \chi^2(\mathcal{Z}_{0d_{3/2}})
  &=
  \sum_i
  \frac{
  \left(\mathcal{Z}_{0d_{3/2}} \sigma_i^{\textrm{DWIA}}-\sigma_i\right)^2
  }{
  \delta_i^2}.
\end{align}
$\sigma_i^{\textrm{DWIA}}$ and $\sigma_i$ are theoretical and experimental 
cross sections at data points, respectively, and $\delta_i$ is associated error of the experimental data.
Obtained spectroscopic factors are summarized in Table~\ref{tab:s-factor}. 
Following Ref.~\cite{Wakasa17}, only the data points around the peak, larger than $25~\mu$b/(MeV sr$^2$),
are fitted to reduce the uncertainty.
\begin{table}[bp]
\caption{\label{tab:s-factor}
Setup and resulting spectroscopic factors.
}
\begin{ruledtabular}
\begin{tabular}{lllc}
SPWF   & Optical pot. & $p$-$p$ int. & $\mathcal{Z}_{0d_{3/2}}$ \\
\colrule
Kramer & KD           & FL           & $0.623   \pm 0.006$    \\
Kramer & Dirac        & FL           & $0.672   \pm 0.006$    \\
DOM    & DOM          & FL           & $0.560   \pm 0.005$    \\
DOM    & DOM          & Mel          & $0.489   \pm 0.005$    \\
DOM    & DOM          & Mel (free)   & $0.515   \pm 0.005$    \\
\end{tabular}
\end{ruledtabular}
\end{table}

The spectroscopic factors obtained from the phenomenological ($p$,$2p$) analysis
are consistent with the phenomenological ($e$,$e'p$) analysis which gave
$0.65 \pm 0.06$~\cite{Mack2018}.
On the other hand, the spectroscopic factor obtained using the DOM wave functions to reproduce the ($p$,$2p$) cross section is in disagreement with the DOM-calculated [using Eq.~\eqref{eq:sf}] value of $0.71 \pm 0.04$.
One of the reasons for this inconsistency may lie in a difference in the peripherality 
of the reaction probes, but it is not yet well understood.

The importance of having to deal with three distorted proton waves in the ($p$,$2p$) reaction as compared to just one in the ($e$,$e'p$) case remains an issue. There is an uncertainty associated with the DOM distorted waves due to the experimental data points used in the DOM fit. Considering the strong correlation between the proton reaction cross section and the $^{48}$Ca$(e,e'p)^{47}$K cross section demonstrated in Ref.~\cite{Mack2019}, we look to uncertainties in the experimental proton reaction cross section data points in energy regions corresponding to those of the distorted proton waves to get a rough estimate of the uncertainty associated with the DOM distorted waves. The proton reaction cross-section data points from Refs.~\cite{ca40react-1,ca40react-2} suggest an uncertainty in the corresponding DOM distorted waves around 3\%. Furthermore, due to the kinematics of the reaction, one of the proton energies is as low as 36 MeV. In the DOM analysis of $^{40}$Ca$(e,e'p)^{39}$K, the description of the experimental cross section for outgoing proton energies of 70 MeV, the lowest of the considered proton energies, is unsatisfactory~\cite{Mack2018}. This indicates that the impulse approximation may not be applicable at proton energies of 70 MeV and below. Since one of the outgoing proton energies in this $^{40}$Ca$(p,2p)^{39}$K reaction is even less than 70 MeV, it is reasonable to expect some discrepancy in the $^{40}$Ca$(p,2p)^{39}$K TDX. 
This discrepancy may be reduced when higher proton beam energies are considered but this implies that the DOM analysis has to be extended to higher energies.

\begin{figure}[htbp]
\centering
\includegraphics[width=0.37\textwidth]{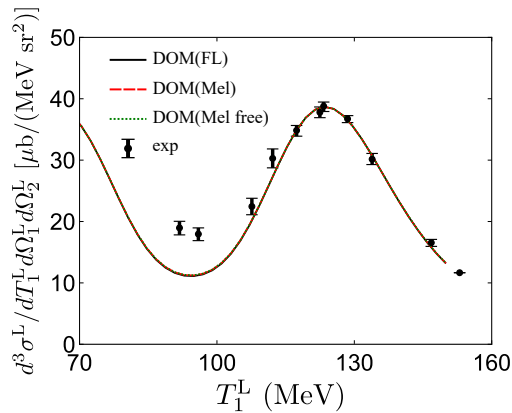}
\caption{
TDX with different $p$-$p$ effective interactions.
The solid, dashed, and dotted lines are TDXs with the 
Franey-Love effective interaction~\cite{Franey85} (FL), 
Melbourne $g$-matrix interaction at mean density~\cite{Amos00} (Mel),
and that at zero density (Mel free),
respectively.
The experimental data are the same as in Fig.~\ref{fig_tdx_dw}.
}
\label{fig_tdx_ppint}
\end{figure}

As a further step to clarify this issue,
we also investigated the uncertainty arising from a 
different choice of the $p$-$p$ effective interactions.
Three different types of the $p$-$p$ effective interactions,
the Franey-Love effective interaction (FL)~\cite{Franey85},
the Melbourne $g$-matrix interaction at mean density (Mel)~\cite{Amos00},
and that at zero density  (Mel free) were utilized.
The mean density of the reaction is defined in Sec. 6.1. of Ref.~\cite{Wakasa17}.
The SPWF and the distorted waves from the DOM
are adopted in this analysis.
The choice of the $p$-$p$ effective interaction only 
changes the magnitude of the TDX, keeping their shapes unchanged
as shown in Fig.~\ref{fig_tdx_ppint}.
Extracted spectroscopic factors are summarized in Table.~\ref{tab:s-factor}.
An uncertainty of the choice of the $p$-$p$ effective interaction 
results in $\mathcal{Z}_{0d_{3/2}} = 0.489$--$0.560$ and does not explain
the inconsistency between the DOM ($p$,$2p$) result and the successful DOM description of the ($e$,$e'p$) cross section~\cite{Mack2018}.

Validity of the finite-range treatment of $t_{pp}$ and factorization approximation as well as
the nonlocality correction and the relativistic correction in nonrelativistic DWIA have 
been discussed for decades~\cite{Jackson67,Austern78,Jackson82,Kanayama90}.
They make certain change in the cross section, but it is still not quantitatively conclusive.
In Ref.~\cite{Yoshida16}, it is reported that TDX 
with AMA only differs 6\% of that with LSCA
in $^{120}$Sn($p$,$p\alpha$)$^{116}$Cd reaction case.
Considering this situation,
it is very important that we obtained good agreement in spectroscopic factors
deduced from phenomenological ($e$,$e'p$) and ($p$,$2p$) analyses 
before discussing DOM + DWIA analysis.
It should be noted that the finite-range treatment of $t_{pp}$ is more crucial in the relativistic DWIA
framework, as pointed out in Ref.~\cite{Ikebata95}.

Ideally, the spectroscopic factor should not depend on the reactions to be adopted. As shown in Ref.~\cite{Wakasa17}, the spectroscopic factors for the SP levels near the Fermi energies of stable nuclei extracted from ($p,2p$) reactions above 200~MeV are consistent with those from ($e,e'p$) with uncertainties ranging from 10\% to 15\%. It should be noted that the nonlocality correction to the SPWF and distorted waves are considered to be a primary source of the uncertainties of the spectroscopic factor~\cite{Wakasa17}. As mentioned above, $\mathcal{Z}_{0d_{3/2}}$ obtained with the DOM-DWIA analysis of the $^{40}$Ca($p,2p$) data at 200~MeV, in which the nonlocality is treated in a sophisticated manner,  differs  by
at least 21\%
from the value used to reproduce $(e,e'p)$ data using the same SPWF and proton distorted wave calculated with the DOM. This implies the necessity of improving the treatment of $p$-$p$ scattering inside a nucleus beyond the standard $t$- or $g$-matrix approach.
One immediate concern is that present treatments of the effective interaction do not allow for energy transfer in the elementary process.
Since in the ($p,2p$) reaction a substantial excitation energy is involved, it implies that the mediators of the strong interaction, in particular the pion, must be allowed to propagate in the system and are certainly not static~\cite{Jonathan11}.

The ability of the DOM to provide both bound and scattering nucleon wave functions is opening a door to a new research opportunity for the nucleon-nucleon scattering process in many-body systems.
This is of particular importance as the ($p,2p$) reaction can be employed in inverse kinematics~\cite{Atar:2018,pty011}. 
There is therefore a clear need to pursue an improved description of the effective interaction in the medium which will also depend on the nucleon asymmetry that is studied in exotic systems.

\section{Summary}
\label{sec_summary}
The DOM SPWF and distorted waves have been applied to the DWIA analysis of ($p$,$2p$) reaction for the first time.
The proton $0d_{3/2}$ spectroscopic factor $\mathcal{Z}_{0d_{3/2}} = 0.56 \pm 0.005$ of $^{40}$Ca was 
extracted from $^{40}$Ca($p$,$2p$)$^{39}$K analysis with the present framework.
The obtained spectroscopic factor is inconsistent with the DOM supplied one that
reproduces $^{40}$Ca($e$,$e'p$)$^{39}$K data, $0.71 \pm 0.04$.
We tested several types of input for the $p$-$p$ effective interaction: 
the Franey-Love interaction, the Melbourne $g$-matrix at mean density, and that at zero density.
In addition to using the DOM wave functions, Kramer's SPWF and
distorted waves obtained by the Koning-Delaroche and Dirac phenomenology optical potentials 
were considered.
We conclude from the present analysis that
the gap between the above-mentioned ($p$,$2p$) and ($e$,$e'p$) spectroscopic factors 
cannot be explained by the uncertainties arising from these inputs.
This result implies there is room for improving 
the treatment of the $p$-$p$ binary scattering beyond the traditional 
$t$- and $g$-matrix approach combined with the impulse approximation.

\section*{ACKNOWLEDGMENTS}
The authors thank T. Noro for providing them with updated
experimental data before publication.
This work was supported by JSPS KAKENHI Grant No. JP20K14475. This work was also supported by the U.S. National Science Foundation under Grant No. PHY-1912643.
TRIUMF receives federal funding via a contribution agreement with the National Research Council of Canada.

\appendix*
\section{Factorization of $T$ matrix}
\label{appendixA}
\begin{figure}[htbp]
\centering
\includegraphics[width=0.30\textwidth]{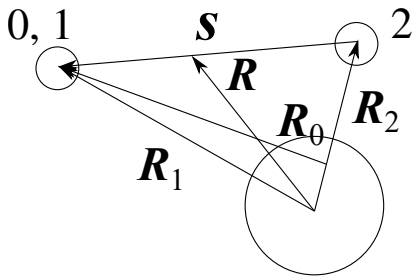}
\caption{
Coordinate of the ($p$,$2p$) three-body system.
}
\label{fig_coordinate}
\end{figure}

In this appendix, the factorization procedure of the $T$ matrix
is explained
by means of the LSCA, following previous research~\cite{Luo91,Watanabe99,Ogata07,Minomo10}.
Equation~(\ref{eq:tmatrix}) is originally an integral over two 
coordinates $\bm{R}$ and $\bm{s}$:
\begin{widetext}
\begin{align}
T_{\mu_1 \mu_2 \mu_0 \mu_j}
&=
\sum_{\mu'_1 \mu'_2 \mu'_0 \mu_p}
\int d\bm{R}\,d\bm{s}\,
\chi_{1, \mu'_1 \mu_1}^{(-)*}(\bm{R}_1)
\chi_{2, \mu'_2 \mu_2}^{(-)*}(\bm{R}_2)
t_{pp}(\bm{s})
[1-P]
\chi_{0, \mu'_0 \mu_0}^{(+)}(\bm{R}_0)
\sum_{m}
\left(lms_p\mu_p | j\mu_j\right)
\psi^n_{ljm}(\bm{R}_2).
\label{eq_originaltmat}
\end{align}
\end{widetext}
$\bm{R}$, $\bm{s}$, $\bm{R}_0$, $\bm{R}_1$, and $\bm{R}_2$ 
are defined as shown in Fig.~\ref{fig_coordinate}.
$t_{pp}(\bm{s})$ is a $p$-$p$ effective interaction and $P$ is the exchange operator for the colliding two protons; thus, as usual, both the direct and exchange terms are considered.
$\bm{R}$, $\bm{s}$, $\bm{R}_0$, $\bm{R}_1$, and $\bm{R}_2$ 
are rewritten in terms of $\bm{R}$ and $\bm{s}$ as
\begin{align}
\bm{R}_0
&=
\left(1-\alpha_{R}\right)\bm{R} + \frac{A+1}{2A}\bm{s},\\
\bm{R}_1
&=
\bm{R}+\frac{1}{2}\bm{s},\\
\bm{R}_2
&=
\bm{R}-\frac{1}{2}\bm{s}.
\end{align}
The nucleon and the target mass number are 1 and $A$, respectively.
The factorization of the $p$-$p$ $t$-matrix is done by considering three
approximations as follows.
First, for a short distance $\Delta \bm{R}$, the local semiclassical 
approximation (LSCA)~\cite{Luo91,Watanabe99,Ogata07,Minomo10}
can be applied to the distorted waves:
\begin{align}
\chi(\bm{R}+\Delta \bm{R})
&\approx
\chi(\bm{R})e^{i\bm{K}(\bm{R})\cdot \Delta\bm{R}}.
\end{align}
$\bm{K}(\bm{R})$ is the local momentum of the scattering particle at $\bm{R}$.
Further approximation is made  
by replacing the local momentum $\bm{K}(\bm{R})$ with its asymptotic
one $\bm{K}$. This is called the asymptotic momentum approximation (AMA)
In Ref.~\cite{Minomo10}, in which the AMA was referred to as LSCA-A, the AMA was shown to work with almost the same accuracy as of LSCA for a short propagation of the distorted waves of about 1.5~fm. It should be noted that for the factorization of the $p$-$p$ transition matrix, the propagation range of about the half the range of $t_{pp}(\bm{s})$ needs to be approximated. According to the result of Ref.~\cite{Minomo10}, both the LSCA and AMA are expected to be sufficiently accurate for this purpose. Furthermore, it is reported in Ref.~\cite{Yoshida16} that 
the knockout cross section with AMA differs only $\sim6\%$ from that with LSCA
in $^{120}$Sn($p$,$p\alpha$)$^{116}$Cd reaction at 392~MeV.

Thus, the AMA (and the LSCA as well) can safely be applied to the distorted waves in Eq.~(\ref{eq_originaltmat}), which reads
\begin{align}
\label{eq_ama}
\chi_{0,\mu'_0\mu_0}^{(+)}(\bm{R}_0)
\approx&
\chi_{0,\mu'_0\mu_0}^{(+)}(\bm{R}) 
e^{-\alpha_R\bm{K}_0\cdot\bm{R}}\nonumber \\
&\times e^{i(A+1)\bm{K}_0\cdot \bm{s}/2A},\\
\chi_{1,\mu'_1\mu_1}^{(-)*}(\bm{R}_1)
\approx&
\chi_{1,\mu'_1\mu_1}^{(-)*}(\bm{R})
e^{i\bm{K}_1\cdot \bm{s}/2},\\
\chi_{2,\mu'_2\mu_2}^{(-)*}(\bm{R}_2)
\approx&
\chi_{2,\mu'_2\mu_2}^{(-)*}(\bm{R})
e^{-i\bm{K}_2\cdot \bm{s}/2},
\end{align}
with $\bm{K}_i$ ($i = 0, 1, 2$) being the asymptotic momenta of particle $i$. Note that in Eq.~(\ref{eq_ama}), we have also used the fact that $\alpha_{R}$ is small.
As for the single-particle wave function, the Fourier transformation
is applied to factorize out the $\bm{s}$ dependence
\begin{align}
\psi^n_{ljm}(\bm{R}_2)
&=
\frac{1}{(2\pi)^{3}}
\int d\bm{K}_N\, \tilde{\psi}^n_{ljm}(\bm{K}_N)e^{i\bm{K}_N \cdot \bm{R}_2} \nonumber \\
&=
\frac{1}{(2\pi)^{3}}
\int d\bm{K}_N\, \tilde{\psi}^n_{ljm}(\bm{K}_N)e^{i\bm{K}_N \cdot (\bm{R}-\bm{s}/2)}.
\label{eq_fourier}
\end{align}

The second approximation is to assume the quasi-free $p$-$p$ collision 
in the ($p$,$2p$) reaction. 
Although the distorted waves are not eigenstates of momenta,
the initial (final) $p$-$p$ relative momentum $\bm{\kappa}$ ($\bm{\kappa}'$)
relevant to the ($p$,$2p$) reaction are approximately given by the asymptotic momenta $\bm{K}_i$ as
\begin{align}
\bm{\kappa} 
&\approx
\frac{A+1}{2A}\bm{K}_0-\frac{1}{2}\bm{K}_N \\
\bm{\kappa}' 
&\approx
\frac{1}{2}\bm{K}_1-\frac{1}{2}\bm{K}_2.
\end{align}
It should be noted that this is a natural outcome of the use of the AMA mentioned above. 
Similarly, as the last approximation, the momentum conservation of the $p$-$p$ collision is assumed:
\begin{align}
\frac{A+1}{A}\bm{K}_0 + \bm{K}_N 
\approx
\bm{K}_1 + \bm{K}_2.
\label{ppmomcons}
\end{align}
Again, the appearance of the asymptotic momenta in Eq.~(\ref{ppmomcons}) is due to the use of the AMA.

Under these approximations,
once the ($p$,$2p$) kinematics is fixed, 
$\bm{K}_N$ and $\bm{\kappa}$ are uniquely determined by $\bm{K}_0$, $\bm{K}_1$ and $\bm{K}_2$.
Inserting Eqs.~(\ref{eq_ama})--(\ref{eq_fourier}) into Eq.~(\ref{eq_originaltmat})
and applying the inverse Fourier transform of $\tilde{\psi}_{nlm}^{n}(\bm{K}_N)$,
one obtains the factorized form of the $T$ matrix
\begin{widetext}
\begin{align}
T_{\mu_1 \mu_2 \mu_0 \mu_j}
=
&\sum_{\mu'_1 \mu'_2 \mu'_0 \mu_p}
\int d\bm{s}\,
e^{-i\bm{\kappa}'\cdot\bm{s}}
t_{pp}(\bm{s})
[1 - P]
e^{i\bm{\kappa}\cdot\bm{s}}
\int d\bm{R}\,
\chi_{1, \mu'_1 \mu_1}^{(-)*}(\bm{R})
\chi_{2, \mu'_2 \mu_2}^{(-)*}(\bm{R})
\chi_{0, \mu'_0 \mu_0}^{(+)}(\bm{R})
e^{-i\alpha_\mathrm{R}\bm{K}_0\cdot\bm{R}}\sum_{m}
\left(lms_p\mu_p | j\mu_j\right)
\psi^n_{ljm}(\bm{R}).
\end{align}
\end{widetext}
This is equivalent to Eqs.~(\ref{eq:tmatrix}) and (\ref{eq:tpp}), if one takes account of the spin degrees of freedom in the $p$-$p$ $t$ matrix.

\bibliography{ref}

\end{document}